Fabrication of small superconducting coils using (Ba,$A$)Fe$_2$As$_2$ ($A$: Na, K) round wires with large critical current densities

Sunseng Pyon[1], Haruto Mori[1], Tsuyoshi Tamegai[1], Satoshi Awaji[2], Hijiri Kito[3], Shigeyuki Ishida[3], Yoshiyuki Yoshida[3], Hideki Kajitani[4], Norikiyo Koizumi[4]

[1]Department of Applied Physics, The University of Tokyo, Bunkyo, Tokyo 113-8656, Japan
[2]High Field Laboratory for Superconducting Materials, Institute for Materials Research, Tohoku University, Sendai 980-8577, Japan
[3]Research Institute for Advanced Electronics and Photonics, National Institute of Advanced Industrial Science and Technology, Tsukuba, Ibaraki 305-8568, Japan
[4]Naka Fusion Institute, National Institutes for Quantum and Radiological Science and Technology (QST), 801-1 Mukoyama, Naka-shi, Ibaraki 311-0193, Japan

**Abstract**
We report the fabrication of small (Ba,$A$)Fe$_2$As$_2$ ($A$: Na, K) coils using 10 m-class long round wires, fabricated by powder-in-tube (PIT) method. Coils are sintered using hot-isostatic-press (HIP) technique after glass-fiber insulations are installed. Critical current ($I_c$) of the whole coil using (Ba,Na)Fe$_2$As$_2$ and (Ba,K)Fe$_2$As$_2$ are 60 A and 66 A under the self-field, and the generated magnetic fields at the center of the coil reach 2.6 kOe and 2.5 kOe, respectively. Furthermore, the largest transport critical current density ($J_c$) and $I_c$ in (Ba,Na)Fe$_2$As$_2$ wires picked up from the coil reach 54 kAcm$^{-2}$ and 51.8 A at $T$ = 4.2 K under a magnetic field of 100 kOe, respectively. This value exceeds transport $J_c$ of all previous iron-based superconducting round wires. Texturing of grains in the core of the wire due to the improvement of the wire drawing process plays a key role for the enhancement of $J_c$.

**Introduction**

Iron-based superconductors (IBSs) are expected to be the next generation high temperature superconductors for high magnetic field applications. Among IBSs, 122-type compounds such as $(AE,A)Fe_2As_2$ ($AE$ = Ba, Sr, $A$ = Na, K) are considered to be promising materials for practical use. Both critical temperature $T_c$ (< 38 K) [1-6] and upper critical field $H_{c2}$ (> 650 kOe) [2,7-9] of $(AE,A)Fe_2As_2$ are larger than those of practically used NbTi and Nb$_3$Sn. Compare with MgB$_2$ with similar $T_c$ (~39 K) [10], which is developed to be used for MRI, $H_{c2}$ is much larger. Small anisotropy $\gamma$ < 2 [7-9] and large critical grain boundary angle of ~9º for critical current density ($J_c$) in $(AE,A)Fe_2As_2$, which is larger than that of ~5º in cuprate YBa$_2$Cu$_3$O$_{7-\delta}$ [11], suggest that high and three-dimensional texturing is not necessary unlike tapes and coated conductors of cuprates. So the IBSs are suitable for applications such as superconducting wires, tapes, and coated conductors. Furthermore, robustness of $J_c$ under high magnetic fields in IBSs is also preferred for high-field applications. In $(AE,K)Fe_2As_2$ single crystal, large $J_c$ above 1 MAcm$^{-2}$ below 5 K at self-field has been demonstrated. Furthermore, values of $J_c$ can be increased by adding pinning center by ion-irradiation [12-16]. Superconducting wires and tapes fabricated by powder-in-tube (PIT) method using polycrystalline $(AE,K)Fe_2As_2$ powders have been also studied [17-24]. Transport $J_c$ of wires and tapes can be increased by purification of raw materials [25] and densification of the core [26] to eliminate weak links between superconducting grains. To enhance $J_c$ by densification of superconducting core, pressing at high temperature is effective for IBS wires. Uniaxial hot press technique has been applied for textured tapes [26-36]. On the other hand, hot isostatic press (HIP) methods has been applied for round wires [15,37-49]. The transport $J_c$ under a high field of 100 kOe at 4.2 K for tapes of (Ba,K)Fe$_2$As$_2$ has reached 150 kAcm$^{-2}$, which exceeds the practical level of 100 kA cm$^{-2}$ [31]. In the case of tapes, transport $J_c$ is significantly enhanced by texturing of the grains due to their flat shapes helped by uniaxial pressing [26,44]. In the case of round wires, which are suitable for winding coils, texturing of the grains is difficult compared with that in tapes. However, it has been demonstrated that the degree of texturing of the grains can be enhanced to a certain degree by changing deformation process, leading to larger $J_c$ [44]. The transport $J_c$ under a high field of 100 kOe at 4.2 K for wires of (Ba,K)Fe$_2$As$_2$ and (Ba,Na)Fe$_2$As$_2$ have reached at levels of 38 and 44 kAcm$^{-2}$, respectively.

For practical applications of these promising wires and tapes, further important progresses, such as fabrication of wires with multifilament core [28,50,51] and superconducting joints [52,53], have been reported. One of the most typical applications of superconducting wires and tapes is the fabrication of coils to generate magnetic fields. In 2017, fabrications of 100 m-class (Sr,K)Fe$_2$As$_2$ tapes and coils using these tapes were reported for the first time, although generated magnetic field was not reported [54]. Recently, using (Ba,K)Fe$_2$As$_2$ tapes, fabrications of several coils and evaluation of critical current $I_c$ have been reported [55-57]. For example, the transport $I_c$ of the pancake coil achieved 35 and 25 A under magnetic fields of 100 and 240 kOe at 4.2 K, respectively [55]. A racetrack coil was also fabricated, and it quenched at 100 kOe at 4.2 K near the operating current of 65 A, which is as large as 86.7% of $I_c$ of the short sample at the same condition [57]. These results clearly indicate that IBSs are very promising for high-field magnet applications. On the other hand, there have been few reports on IBS round wires longer than 1 m or a coil using round wires. Furthermore, not much attention has been paid to $I_c$ of round wires, although there have been several reports on $J_c$ of the wire. Very recently, relatively large $I_c$ in short pieces of round wires, and the fabrication of a small demo coil using round wires with length up to 9 m has been reported [58]. $I_c$ in a short piece of the wire under the magnetic field of 10 or 100 kOe at 4.2 K are 95 or 54 A, respectively. $I_c \sim 54$ A at 100 kOe corresponds to $J_c \sim 42$ kAcm$^{-2}$, which is comparable to the largest value of $J_c$ in round wires (44 kAcm$^{-2}$) in previous reports [48,49]. A small demo coil, fabricated without insulation, generated 1.5 kOe at 60 A at 4.2 K, which was smaller than the expected value [58]. In addition, the magnetic field shows appreciable time dependence with a long time constant due to no insulation. These results suggest that there is a lot of room for improvements of coil fabrications with IBS round wires. In the case of MgB$_2$, coils using superconducting PIT wire have been studied [59,60]. Using a 58 m-long PIT wire, a superconducting magnet was fabricated, which generated a magnetic field of 19 kOe at 4.2 K [60]. Demonstration of generating magnetic field using IBS coil is also demanded.

In this report, we demonstrate the fabrication of small coils using 10 m-class long round wires of (Ba,Na)Fe$_2$As$_2$ and (Ba,K)Fe$_2$As$_2$. Round wires are fabricated by PIT method. Coils are wound by using long wires covered by insulating fiber-glass sleeves, then sintered using HIP technique. Generated magnetic field of the coils and their $I_c$ are also evaluated. Furthermore, largest transport $J_c$ in the short segment picked up from the (Ba,Na)Fe$_2$As$_2$ coil reached 54 kAcm$^{-2}$ at $T = 4.2$ K under a magnetic field of 100 kOe. This value exceeds the value of transport $J_c$ of all previous IBS round wires. Details of the fabrication and characterization of wires and coils, and characterizations of short segments of the wire picked up from the coils including X-ray diffraction (XRD) and X-ray computed tomography (CT) are shown and discussed.

**Experimental methods**

(Ba,Na)Fe$_2$As$_2$ and (Ba,K)Fe$_2$As$_2$ round wires were fabricated by *ex situ* PIT method. Polycrystalline powders of (Ba,Na)Fe$_2$As$_2$ and (Ba,K)Fe$_2$As$_2$ were synthesized by solid state reaction after mixing the raw materials of elements by ball milling method. Details of synthesis are described in Ref. [48] and Ref. [42]. It should be noted that raw materials of (Ba,K)Fe$_2$As$_2$ before synthesis were accidently exposed to a little amount of air, which may affect the results on (Ba,K)Fe$_2$As$_2$ wire and coil as described below. After grinding by agate motor in Ar atmosphere, polycrystalline power was filled into a Ag tube with outer diameter (OD) and inner diameter (ID) of 4.5 mm and 3 mm, respectively. The Ag tube filled with the polycrystalline powder was cold-drawn successively using dies with circular holes down to a diameter of ~1.9 mm. Obtained wires were inserted into Cu tubes with OD and ID of 3.0 mm and 2.0 mm, respectively. It should be noted that we used Cu tubes with different outer and inner diameters compared with previous reports (OD and ID of 3.2 mm and 1.6 mm) to increase $I_c$ of the wire by enhancing the cross-sectional core area [44,48]. They were cold-drawn using dies with circular holes, and formed into a round shape with a diameter of ~2.3 mm (~2.6 mm) for $(Ba,Na)Fe_2As_2$ ($(Ba,K)Fe_2As_2$). Then, the tube was swaged using a rotary swaging machine, and formed into a round shape with a diameter of 1.0 mm (1.16 mm) for $(Ba,Na)Fe_2As_2$ ($(Ba,K)Fe_2As_2$). The total lengths of long wires reach ~12.5 m and ~10 m for $(Ba,Na)Fe_2As_2$ and $(Ba,K)Fe_2As_2$ wires, respectively. For insulation of the long wires, fiber-glass sleeves were installed, which increased the total diameter of the wire to ~1.3 mm. Then, they were wound around stainless steel bobbins, and fixed by being tied up with fiber-glass sleeves. After sealing the edges of wires by an arc welder, coils were sintered at 700°C for 4 h using the HIP technique. HIP processes for $(Ba,Na)Fe_2As_2$ and $(Ba,K)Fe_2As_2$ coils were performed under 200 MPa at National Institute of Advanced Industrial Science and Technology and under 190 MPa at National Institutes for Quantum and Radiological Science and Technology, respectively. The $I_c$ of the coils under the self-field and generated magnetic field at the center of coils were measured at The University of Tokyo. In order to minimize the effect of Joule heating at the current leads, measurements were performed in liquid helium. Current–voltage (*I–V*) characteristics were measured by the four-probe method with solder for contacts, which were made at ends of the wires for coils. The generated magnetic field at the center of the coil was measured using a Hall probe (THS126, Toshiba). After evaluation of properties of coils, short segments of wires were picked up from coils at every 1 m distance from the end. $I_c$ of the short segments of the wire from $(Ba,Na)Fe_2As_2$ coil were measured in static magnetic field up to 140 kOe using the 15T-SM at the High Field Laboratory for Superconducting Materials, IMR, Tohoku University. Similar measurements on short segments of the wire from $(Ba,K)Fe_2As_2$ coil under the self-filed were performed at The University of Tokyo. For the X-ray CT imaging of short segments of wires from $(Ba,Na)Fe_2As_2$ and $(Ba,K)Fe_2As_2$ coils, we used Carl Zeiss METROTOM 1500 with setting 200 kV for X-ray tube voltage or SMX-160CT-SV3 (Shimadzu corporation) with setting 160 kV for X-ray tube voltage, respectively. The bulk magnetization of a short piece of the wire was measured to characterize the superconducting transition and magnetic $J_c$ by a superconducting quantum interference device magnetometer (MPMS-5XL, Quantum Design). Vickers hardness, *HV* [61], was measured on the polished surface of the wire core. Powder XRD with Cu-K$\alpha$ radiation (Smartlab, Rigaku) for the polycrystalline powders and the core of the wires were carried out for the evaluation of texturing of the core of the wires.

**Experimental results and discussion**

Photos of small superconducting coils fabricated from $(Ba,Na)Fe_2As_2$ and $(Ba,K)Fe_2As_2$ wires are shown in Figs. 1(a) and (b), respectively. Specifications of wires and coils including geometrical dimensions and $I_c$ are summarized in Table 1. The *I-V* characteristics and current dependence of generated magnetic field of the $(Ba,Na)Fe_2As_2$ and $(Ba,K)Fe_2As_2$ coils are shown in Figs. 2(a) and 2(b), respectively. As shown in Fig. 2(a), when the current is smaller than ~60 A, a linear increase of the voltage is observed, which amounts to the resistance of ~1 mΩ in the $(Ba,Na)Fe_2As_2$ coil. This finite resistance originates from the resistance of Cu wire without superconducting core at ends of the wire with a total length of ~10 cm, which we found out later. We evaluated the transport $I_c$ for coils by adopting the 1 µVcm$^{-1}$ criterion after subtract the linear component originated from the resistance of Cu. Then $I_c$ of the $(Ba,Na)Fe_2As_2$ coil can be estimated as ~60 A at 4.2 K under the self-field. The generated magnetic field increases linearly as a function of the current both below and above $I_c$ as expected. This indicates that insulation of the wound wire works effectively, in contrast to the coil without insulation in the previous report [58]. At 60 A, generated magnetic field reaches 2.6 kOe, which corresponds to the coil constant of the $(Ba,Na)Fe_2As_2$ coil of ~43 Oe/A. Similarly, as shown in Fig. 2(b), the $(Ba,K)Fe_2As_2$ coil generated 2.5 kOe at $I_c$ of 66 A, which corresponds to the coil constant of ~38 Oe/A. A linear increase of the voltage is also observed in the $(Ba,K)Fe_2As_2$ coil by the same reason. The magnetic field at the center of a coil with finite dimensions can be calculated using the following formula [62];

$$H = \frac{\mu_0 N I}{2(R_2 - R_1)} \ln \frac{R_2 + \sqrt{R_2^2 + l^2}}{R_1 + \sqrt{R_1^2 + l^2}},$$

where $\mu_0$ is vacuum permeability, *N* is the numbers of turns, 2*l* is the height of the coil along its axis, $2R_1$ and $2R_2$ are ID and OD of the coil, respectively. Values of these parameters are also summarized in Table 1. The calculated coil constants of 41 Oe/A and 39 Oe/A, *H/I*, are close to the measured coil constants of 43 Oe/A and 38 Oe/A for the $(Ba,Na)Fe_2As_2$ and $(Ba,K)Fe_2As_2$ coils, estimated from the slope of *H-I* graphs in Figs. 2(a) and 2(b), respectively. These results indicate that both $(Ba,Na)Fe_2As_2$ and $(Ba,K)Fe_2As_2$ coils are properly fabricated as designed.

After the characterization of generated magnetic field, wires of the coils were unwound and short segments of wires were picked up at every 1 m from the end, and they were evaluated. Figs. 3(a) and 3(b) exhibit $I_c$ of these short segments of wires from $(Ba,Na)Fe_2As_2$ and $(Ba,K)Fe_2As_2$ coils, respectively. We evaluated the transport $I_c$ for short segments of the wires by adopting the 1 µVcm$^{-1}$ criterion. As shown in Fig. 3(a), $I_c$ varies considerably along the length of the $(Ba,Na)Fe_2As_2$ wires. The maximum values of $I_c$ of $(Ba,Na)Fe_2As_2$ wires reach 168, 104, and 51.8 A at 0, 10, and 100 kOe, respectively, for a segment at 4 m from the end of the wire. On the other hand, $I_c$ under the self-field varied between 80 A and 170 A for segments between 0 m and 10 m from the end. Such a large variation of $I_c$ suggests the presence of small cracks or inhomogeneities in the superconducting core in the 10-m class long wire wound in the coil. It should be noted that transport $J_c$ of the IBS wire can be underestimated by accidental presence of micro cracks in the superconducting core. Empirically, values of transport $I_c$ is distributed roughly in the range of 70~100%, when several pieces of wires fabricated in the same process are measured. Furthermore, in the present study, wires were picked up after unwinding from the coil. Obviously, unwinding process can introduce small cracks. Considering these effects, $I_c$ variation can be explained. Furthermore, the segments at 11 m and 12 m from the end shows significantly small $I_c$ below 22 A. This may be caused by accidental stress applied during the unwinding process. These inhomogeneities and partial damage in the wire of the coil cause the relatively smaller $I_c$ of ~60 A of the $(Ba,Na)Fe_2As_2$ coil compared with the largest $I_c$ of ~170 A for short segments. In the case of short segments picked up from the $(Ba,K)Fe_2As_2$ coil, values of $I_c$ varied between 60 A and 110 A. The maximum $I_c$ of short segments picked up from the $(Ba,K)Fe_2As_2$ coil is smaller than that from $(Ba,Na)Fe_2As_2$ coil, although the cross sectional area of superconducting core of $(Ba,K)Fe_2As_2$ is similar to that of $(Ba,K)Fe_2As_2$. However, details inspections of the $(Ba,K)Fe_2As_2$ wire show the presence of inhomogeneities as shown below. The minimum $I_c$ of ~60 A is comparable to the $I_c$ of ~66 A of the $(Ba,K)Fe_2As_2$ coil.

To further characterize the wire, optical micrographs of the cross section and X-ray CT images are taken for short segments of $(Ba,Na)Fe_2As_2$ and $(Ba,K)Fe_2As_2$ wires. Both optical and CT images of the transverse cross section of wires are shown in Figs. 4(a), (d), and Figs. 4(b), (e), respectively, showing nearly circular shapes of wires and isotropic shapes of superconducting cores, which were fabricated by drawing using dies with circular holes and swaging using a rotary swaging machine. Longitudinal cross sections of wires were also observed by X-ray CT as shown in Figs. 4(c) and 4(f). Some degrees of sausaging of the superconducting core can be observed in $(Ba,Na)Fe_2As_2$ wire, while the core is nearly homogeneous in $(Ba,K)Fe_2As_2$ wire. It should be noted that transport $I_c$ are not overestimated due to the fluctuation of the longitudinal cross section area. In principle, the value of transport $I_c$ can be determined by the minimum cross section area of the core between two voltage terminal. The distance between two voltage terminals for measurements of $I_c$ is 5~10 mm. As is observed in the X-ray CT images of Fig. 4, the period of sausaging of the core of the wire is shorter than that distance. We cut the wire 10~20 mm away from the terminal to evaluate the area of cross section of the wire for the evaluation of transport $J_c$. This randomly chosen cross sectional area of the core is always equal to or larger than the minimum area of the wire core. So, at least, the value of transport $J_c$ is not be overestimated. It should be noted that, as shown in Fig. 4(d), numerous black dots can be observed in the superconducting core of the wire picked up from the $(Ba,K)Fe_2As_2$ coil. They are voids in the core or KOH attached on the surface of the core. Existence of KOH suggests two possibilities. First, KOH just exist as an impurity phase in the core. Second, unreacted K in the superconducting core reacted with moisture in the air. These impurities are mainly caused by accidental exposure of raw materials to air for a few seconds as described above. By contrast, such black dots are not observed in the superconducting core of (Ba,Na)Fe$_2$As$_2$ wire as shown in Fig. 4(a). Presence of voids and residual KOH explain the smaller $I_c$ in short segments picked up from the (Ba,K)Fe$_2$As$_2$ coil compared with that from the (Ba,Na)Fe$_2$As$_2$ coil.

After the evaluation of the area of superconducting core of short segments of the wire from (Ba,Na)Fe$_2$As$_2$, it is found that the value of transport $J_c$ at high magnetic field of 100 kOe exceeds the record value of all IBS round wires. The magnetic field dependence of the transport $J_c$ (*I–V*) and magnetic $J_c$ (*M-H*) at 4.2 K for the wires picked up from the (Ba,Na)Fe$_2$As$_2$ coils are shown in Figs. 5(a) and (b), respectively. Transport and magnetic $J_c$ in the (Ba,Na)Fe$_2$As$_2$ HIP wire, which were the largest values of $J_c$ in IBS round wires, from our previous publication are also plotted [49]. As shown in Fig. 5(a), transport $J_c$ under the magnetic field of 10 and 100 kOe of the wires picked up from the (Ba,Na)Fe$_2$As$_2$ coil at 4 m (7 m) reach 109 (114) and 54 (52) kAcm$^{-2}$, respectively. The transport $J_c$ of 54 kAcm$^{-2}$ under 100 kOe at 4.2 K is more than 20% larger than the previous largest value of $J_c$ of 44 kAcm$^{-2}$ [49]. Larger $J_c$ compared with that in the previous report can be observed in the whole magnetic field range as shown in Fig. 5(a). The magnetic $J_c$ of the same wires plotted in Fig. 5(b) were evaluated from the irreversible magnetization using the extended Bean model [12], 20Δ*M*/*a*(1-*a*/3*b*), where Δ*M*(emu/cm$^3$) is $M_{down}$ - $M_{up}$. $M_{up}$ and $M_{down}$ are the magnetization when sweeping the field up and down, respectively, and *a*(cm) and *b*(cm) are the lateral dimensions of the core, approximated by a rectangle (*a* < *b*), keeping the same area of the actual core. Magnetic field is applied parallel to the current flow direction for the wire. Thickness of the sample for the evaluation of magnetic $J_c$ is ~0.5 mm. Fluctuation of the area of cross section is not so large as shown in X-ray CT images of Fig. 4. Furthermore, we also checked both cross sectional areas of the core are comparable. So the fluctuation of cross section area of the cores does not affect the evaluation of magnetic $J_c$. As shown in Fig. 5(b), magnetic $J_c$ under the self-filed and 40 kOe of the wires picked up from the (Ba,Na)Fe$_2$As$_2$ coil at 4 m (7 m) from the end reach 372 (387) and 67 (66) kAcm$^{-2}$, respectively. The magnetic $J_c$ near the self-field are more than twice larger than the transport $J_c$, while $J_c$ values evaluated by two methods are comparable at higher magnetic fields. Compared with the previous data shown in Fig. 5(b), the magnetic $J_c$ of the present wire is larger in the whole magnetic field range. On the other hand, both transport and magnetic $J_c$ of the short segments of the wire picked up from the (Ba,Na)Fe$_2$As$_2$ coil at 11 m from the end are significantly smaller than those of other wires. As described above, this may be caused by accidental stress applied during the unwinding process. On the other hand, quality of short segments cut from (Ba,K)Fe$_2$As$_2$ coil is worse. In Fig. 6, both transport and magnetic $J_c$ of some short segments cut from (Ba,K)Fe$_2$As$_2$ coils are shown. Compared with the previous value in Ref. [44], the highest $J_c$ among all (Ba,K)Fe$_2$As$_2$ round wires, values of both transport and magnetic $J_c$ are about half of the previous values. This inferior performance should be caused by the accidental exposure of raw materials to air as described above. The present study demonstrates that a part of short segments of the wire picked up from the (Ba,Na)Fe$_2$As$_2$ coil shows the transport $J_c$ of 54 kAcm$^{-2}$ at 100 kOe, which is larger than the previous record of IBS round wires of 44 kA cm$^{-2}$ [48,49], as shown in Fig. 5(a). One of possible key factors for the enhancement of $J_c$ is the quality of polycrystalline powder. To confirm this, characteristics of (Ba,Na)Fe$_2$As$_2$ powders used in this work and in the previous work in Ref. [48] are compared. Temperature dependence of normalized magnetization of two kinds of (Ba,Na)Fe$_2$As$_2$ powders are shown in Fig. 7(a). The onset $T_c$ of ~35 K in both powders are almost the same, although the drop of magnetization near $T_c$ in the powder used in the present work is sharper than that in the previous work. On the other hand, $T_c$ and sharpness of the drop of magnetization near $T_c$ are comparable for both wires fabricated in the present work and in the previous report [49], as shown in Fig. 7 (b). In the case of superconducting cores in sintered wires at high pressure, subtle changes of magnetization due to the presence of lower $T_c$ phases can be hidden behind the large shielding effect of the main body. Still a little bit higher quality of polycrystalline powder in the present study helped to enhance transport $J_c$. Next, texturing in the superconducting core may also play a key role for the enhancement of $J_c$, as discussed for IBS round wires [44,48,49]. It is reported that the degree of texturing of the grains in the core is affected by the method how the wire was drawn [44]. In the present study, long wires for the coils were fabricated using dies with circular holes and a rotary swaging machine, resulting in nearly circular cross section. By contrast, the Cu-Ag-sheathed wires in previous reports were drawn into a square shape with a groove roller at the final stage, although Ag-sheathed wires before packing into Cu tube were drawn by using dies with circular holes [44,48,49]. More isotropic circular cross section compared with the rectangular shape may be more advantageous for the texturing of grains. In order to confirm possible texturing in the core of (Ba,Na)Fe$_2$As$_2$ round wires, we performed the XRD measurements. Figure 8(a) shows (002) and (103) peaks from longitudinal cross sections of the wire picked up from the (Ba,Na)Fe$_2$As$_2$ coil at 4 m from the end with transport $J_c$ at 100 kOe of 54 kAcm$^{-2}$. The relative intensity of (002) peaks compared with that of (103), defined by $r = I(002)/I(103)$, for this wire is ~0.36. The values $r$ of other (Ba,Na)Fe$_2$As$_2$ HIP wires have good correlation with transport $J_c$ at 100 kOe. Namely, $r$ values for wires with $J_c$(100 kOe) = 34, 40, and 44 kAcm$^{-2}$ are ~0.20, ~0.27, and ~0.33, respectively [49]. The value of $r$ is a good parameter for concentric texturing of grains in the core around the long axis of the wire, and it should have a positive effect on the enhancement of $J_c$. We also performed a similar measurement for another part of the segment. As shown in Fig. 8 (b), $r$ value of the wire, picked up from the (Ba,Na)Fe$_2$As$_2$ coil at 1 m from the end, is ~0.35, comparable to the value for another segment. This result suggests that the degree of texturing of grains is homogeneous in the whole coil although $I_c$ have some variation due to micro cracks. We also performed the $HV$ measurement for the same wire. The $HV$ of the present wire is 253 [kgfmm$^{-2}$], which is slightly larger than that of 247 [kgfmm$^{-2}$] in previous report [49]. It was reported that high densification of the core of wires or tapes, which can reduce weak links between superconducting grains, can also help to enhance $J_c$ [26]. Although wires in the present and previous studies were sintered at the same pressure of 200 MPa, the thickness of Cu sheath, shapes of the cross section of the wire (rectangle or circle), and the diameter of the wire are different from each other. These differences may affect *HV* of the wires. From these results, we speculate that texturing of the grains mainly affect for enhancement of transport $J_c$ at high magnetic field, while higher quality of powders and higher densification of the core also contribute to that.

In the present study, we demonstrate the fabrication of both $(Ba,Na)Fe_2As_2$ and $(Ba,K)Fe_2As_2$ coils, which generated magnetic field of 2.6 and 2.5 kOe, and have $I_c$ of 60 and 66 A, respectively. Here we discuss strategy for further enhancement of generated magnetic field using IBS coils. Figure 9 shows the load line and $I_c$ of the $(Ba,Na)Fe_2As_2$ coil together with the $I_c$–$H$ curve of the short segment picked up from the coil at 4.2 K. $I_c$ at the intersection of the load line and field dependence of $I_c$ of the short segment is ~140 A, which is more than twice as large as that of the coil. The suppression of $J_c$ of the coil is due to the inhomogeneities and partial damage in the wound wire as discussed above. Improvements of the fabrication process for good uniformity of current carrying capability for the whole length of the wire are demanded. Load line and field dependence of $I_c$ of the short segment in Fig. 9 suggests that increase of $I_c$ and/or enhancement of the coil constant should be needed to generate magnetic field higher than 10 kOe. The area of cross section of the superconducting core was increased by changing the dimensions of Cu sheath compared with the previous reports [44,48,49]. However, as shown in Fig. 4, the fraction of the superconducting core is still small. The diameter of the wire is more than 2.5 times larger than that of the core. There is still much room for expanding the area of the core by reducing thickness of metal sheaths. Coil constants of the fabricated test coil were 38-43 Oe/A as described in Table 1. According to the formula for the generated magnetic field of the coil, coil dimensions $2l$ and $2R_2$ should be increased to generate higher magnetic field. Furthermore, longer wire is necessary. As described above, the $I_c$ at 10 kOe of the short segments of the wires is ~100 A. So the coil constant should be increased larger than 100 Oe/A to generate a magnetic field higher than 10 kOe. As an example, the coil constant becomes ~100 Oe/A, when $N$, $2l$, $2R_1$, $2R_2$, and the diameter of the wire with insulation are 400, 40 mm, 20 mm, 40 mm, and 1 mm respectively. To fabricate this newly designed coil, a ~40 m-long wire should be prepared. Uniform and long round wires with larger cross section area of superconducting core is needed for generating larger magnetic field in the coil.

Further enhancement of $J_c$ in IBS round wire is also expected. One of the most important parameter to increase $J_c$ is the degree of texturing of superconducting grains. In this report, wires are drawn using mostly dies with circular holes, and a rotary swaging machine was used in the final processes. It was suggested in Ref. [63] that drawing wires using dies is preferred than swaging for increasing $J_c$. We used the rotary swaging machine in the final process to avoid breakage of the thin Cu/Ag-sheathed wires. The breakage of the wire using dies with smaller holes can be avoided by inserting annealing process of the wire after several drawing processes. As described above, a higher degree of grain texturing is evaluated in the wire with largest $J_c$ compared with the previous report. By improving the fabrication process, such as drawing wires exclusively using dies, may realize further texturing of the grains in the core. Next, uniform core shape over the entire length of the wire is also important. As shown in X-ray CT images in Figs. 4(c) and 4(f), slight sausaging of the core and unevenness of the interface between the Ag sheath and the core can be detected. Togano *et al*. reported that the smoothness of the interface between the sheath and the core in $(Ba,K)Fe_2As_2$ tape is significantly improved by using Ag-Sn binary alloy sheaths [64]. Using proper metal sheath, which is hard and stable against core material, uniform core shape in the wire can also be realized. Finally, quality of polycrystalline materials is also a key factor to reduce weak links between the grains and increase $J_c$. The condition of synthesis of polycrystalline sample directly affects the sample quality [42,48]. Kametani *et al*. reported that oxygen and water level should be significantly reduced in the synthesis environment [25]. Otherwise, insulating Ba-O network is formed at the grain boundaries. Actually, $(Ba,K)Fe_2As_2$ powder is degraded by accidental exposure to a little amount of air in the synthesis environment, as described above. Deterioration of qualities of $(Ba,Na)Fe_2As_2$ due to the exposure of small amount of oxygen or water is one of possible reasons for the difference of the quality between this work and the previous work as shown in Fig. 7(a). Reduction of chance for exposure to even a small amount of oxygen or water may be effective for avoiding the formation of weak links between superconducting grains.

**Summary**

We demonstrate the fabrication of small $(Ba,A)Fe_2As_2$ (*A*: Na, K) coils using 10 m-class long round wires, using PIT method. Coils are processed by using HIP technique after insulated round wires are wound on the bobbin. $I_c$ of the whole coils using $(Ba,Na)Fe_2As_2$ and $(Ba,K)Fe_2As_2$ wires are 60 A and 66 A under the self-field, and generated magnetic fields at the center of the coil reach 2.6 kOe and 2.5 kOe, respectively. Furthermore, the largest transport $J_c$ and $I_c$ in $(Ba,Na)Fe_2As_2$ wires picked up from the coil reach 54 kAcm$^{-2}$ and 51.8 A at $T$ = 4.2 K under a magnetic field of 100 kOe, respectively. This value exceeds the value of transport $J_c$ of all previous IBS round wires. Texturing of grains in the superconducting core mainly plays an important role for the enhancement of transport $J_c$ at high magnetic field, and higher quality of powders and higher densification of the core also contribute to that. For generating larger magnetic field in the coil, uniform and long round wires with larger cross sectional area of superconducting core is demanded. To increase $J_c$, improvements of fabrication conditions, selection of proper metal sheaths, and reduction of the chance of exposure of core materials to small amount of oxygen or water should be exercised.

**Acknowledgements**

This work was supported by a Grant-in-Aid for Scientific Research (A) (17H01141) from the Japan Society for the Promotion of Science (JSPS). A part of this study was performed at the High Field Laboratory for Superconducting Materials, Institute for Materials Research, Tohoku University (Project No. 20H0023). We are grateful to Dr. Ohtake and Prof. Suzuki in The University of Tokyo for performing X-ray CT imaging of $(Ba,Na)Fe_2As_2$ wires. A part of this work was supported by NIMS microstructural characterization platform (NMCP) as a program of "Nanotechnology Platform" of the Ministry of Education, Culture, Sports, Science and Technology (MEXT), Japan, Grant Number JPMXP09A20NM0131. We are grateful to Dr. Takenouchi in NMCP for performing X-ray CT imaging of $(Ba,K)Fe_2As_2$ wires.

Table 1. Specifications of Cu-Ag-sheathed (Ba,$A$)Fe$_2$As$_2$ ($A$: Na, K) coils.

| Structure | Parameters | (Ba,Na)Fe$_2$As$_2$ | (Ba,K)Fe$_2$As$_2$ |
|---|---|---|---|
| Wires | Diameter | 1.0 mm | 1.16 mm |
| | Length | ~12.5 m | ~10 m |
| | Filament number | Mono core | Mono core |
| Coils | Inner diameter ($2R_1$) | 19 mm | 19 mm |
| | Outer diameter ($2R_2$) | 32 mm | 35 mm |
| | Height ($2l$) | 42 mm | 25.5 mm |
| | Numbers of turns ($N$) | 160 | 114 |
| | Coil constant (calculated) | 41 Oe/A | 39 Oe/A |
| | Critical current ($I_c$) | 60 A | 66 A |
| | Generated magnetic field ($H$) | 2.6 kOe | 2.5 kOe |
| | Coil constant (measured) | 43 Oe/A | 38 Oe/A |

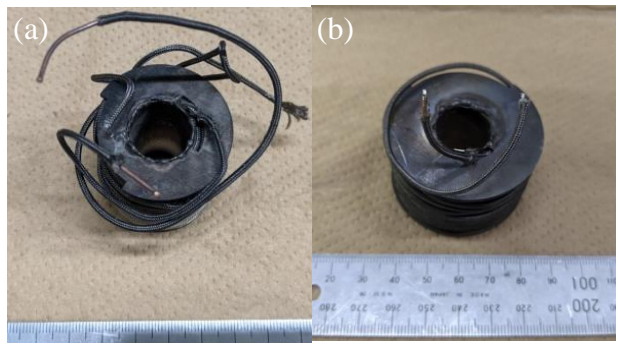

Figure 1. Optical photos of (a) the $(Ba,Na)Fe_2As_2$ coil and (b) the $(Ba,K)Fe_2As_2$ coil.

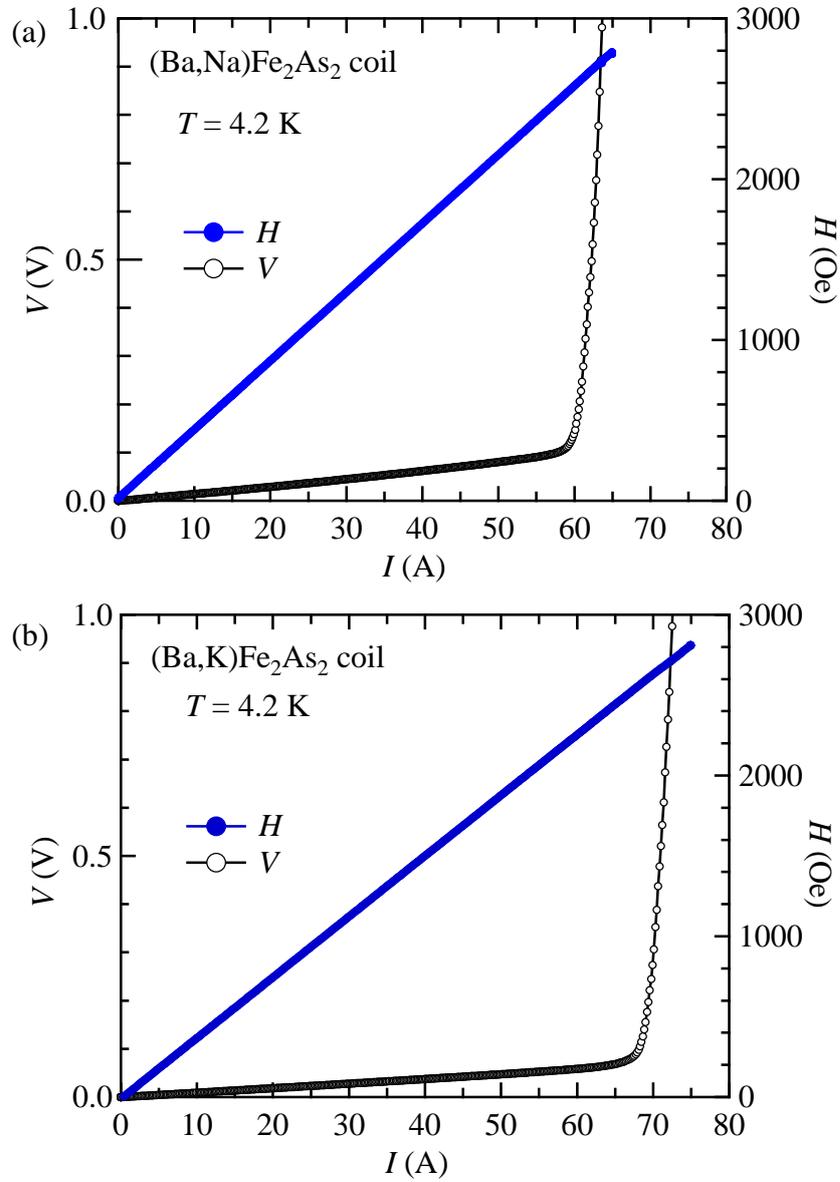

Figure 2. *I-V* characteristics and electric current dependence of generated magnetic field of (a) $(Ba,Na)Fe_2As_2$ and (b) $(Ba,K)Fe_2As_2$ coils. Measurements were performed at 4.2 K under the self-field.

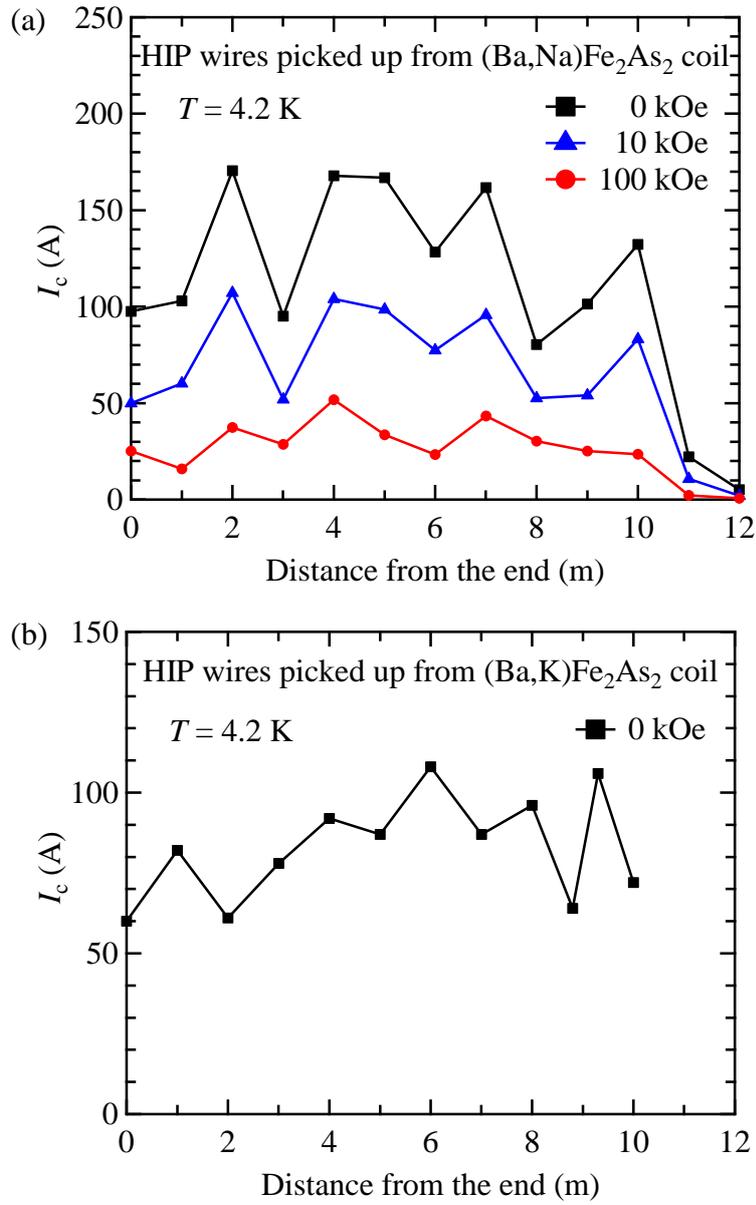

Figure 3. Transport $I_c$ of short segments of wires picked up from (a) (Ba,Na)Fe$_2$As$_2$ and (b) (Ba,K)Fe$_2$As$_2$ coils. Measurements were performed at 4.2 K under several magnetic fields.

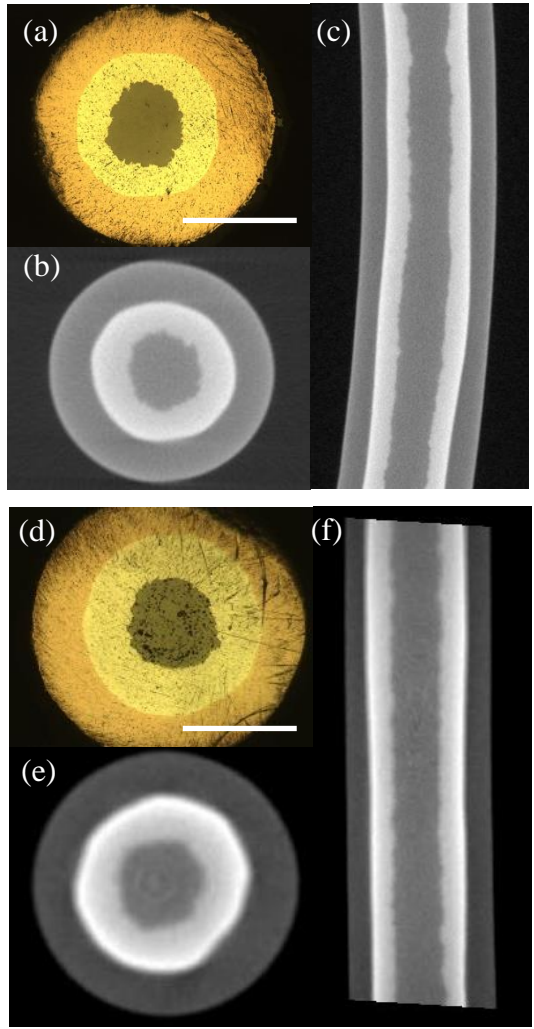

Figure 4. Optical micrographs and X-ray CT images of the transverse cross section of wires picked up from (a)-(b) (Ba,Na)Fe$_2$As$_2$ coil and (d)-(e) (Ba,K)Fe$_2$As$_2$ coil. White lines in (a) and (d) indicate 0.5 mm. X-ray CT images of the longitudinal cross section of wires picked up from (c) (Ba,Na)Fe$_2$As$_2$ and (f) (Ba,K)Fe$_2$As$_2$ coils.

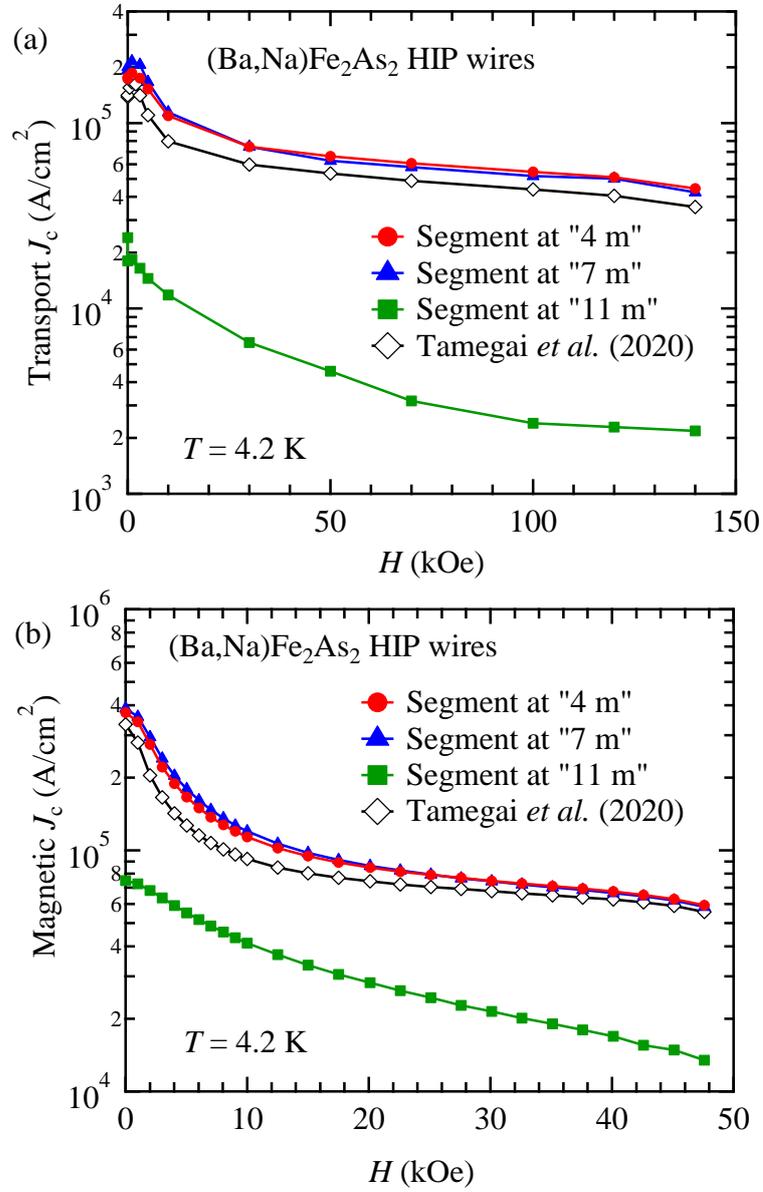

Figure 5. Magnetic field dependence of (a) transport $J_c$ ($I$–$V$) and (b) magnetic $J_c$ ($M$-$H$) at 4.2 K for the short segments picked up from the (Ba,Na)Fe$_2$As$_2$ coil. Transport and magnetic $J_c$ of the (Ba,Na)Fe$_2$As$_2$ HIP wire from our previous study are also plotted for comparison [49].

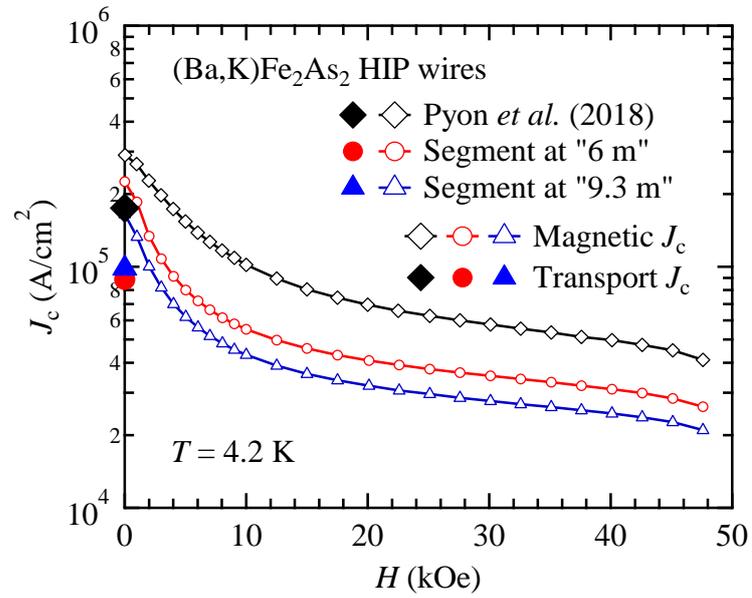

Figure 6. Magnetic field dependence of transport $J_c$ (*I–V*) and magnetic $J_c$ (*M-H*) at 4.2 K for the short segments picked up from the $(Ba,K)Fe_2As_2$ coil. Transport and magnetic $J_c$ of the $(Ba,K)Fe_2As_2$ HIP wire from our previous study are also plotted for comparison [44].

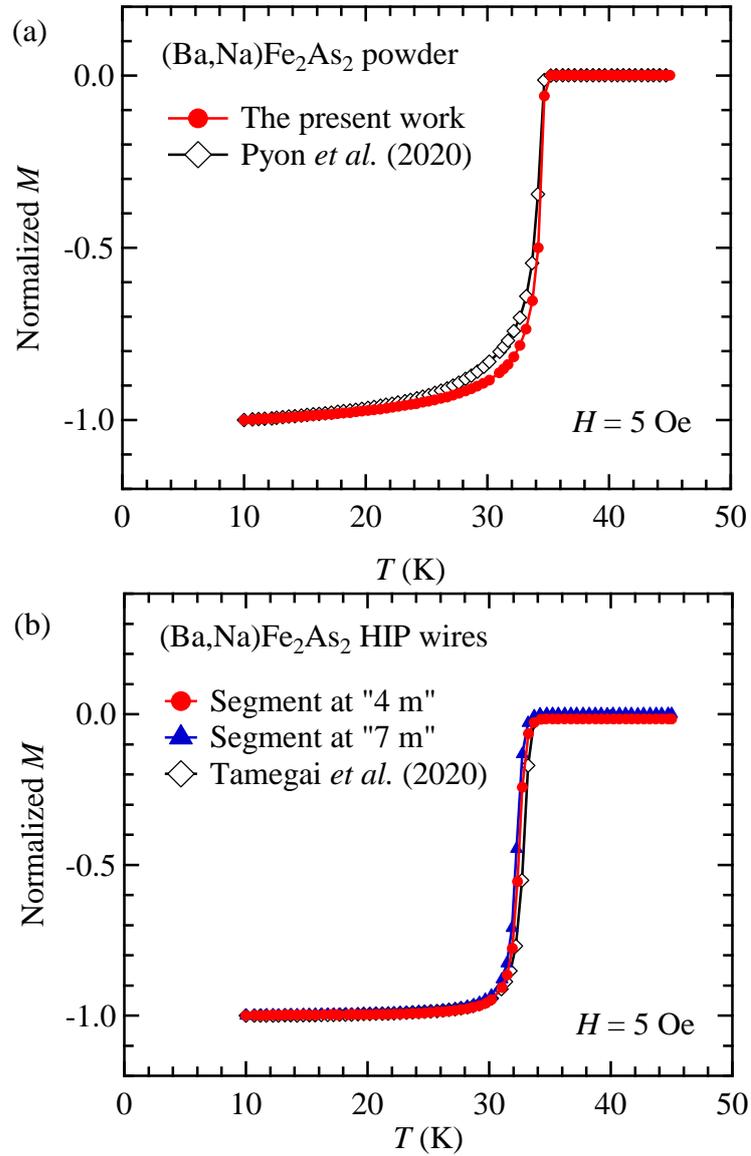

Figure 7. Temperature dependence of normalized magnetization at 5 Oe for (a) $(Ba,Na)Fe_2As_2$ powders and (b) $(Ba,Na)Fe_2As_2$ HIP wires. Data from our previous studies are also plotted for comparison [48,49].

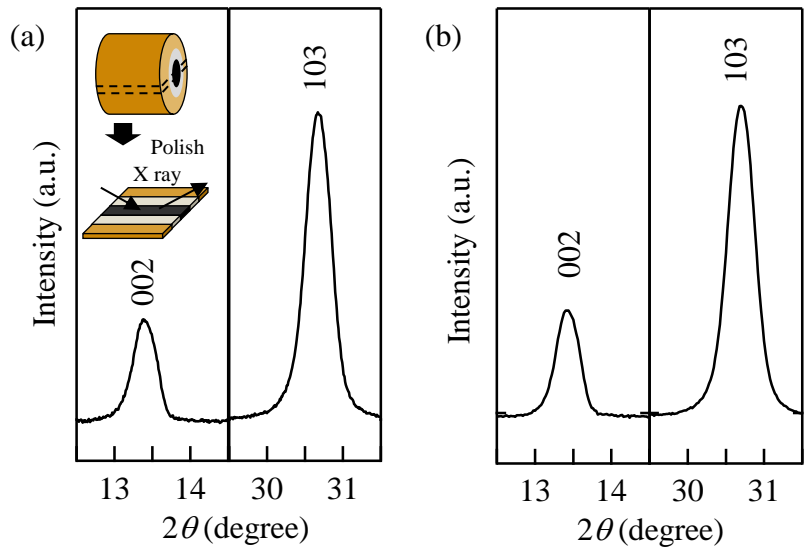

Figure 8. XRD patterns of (002) and (103) peaks for longitudinal cross section of the wire picked up from the (Ba,Na)Fe$_2$As$_2$ coil at a distance (a) 4 m and (b) 1 m from the end of the wire.

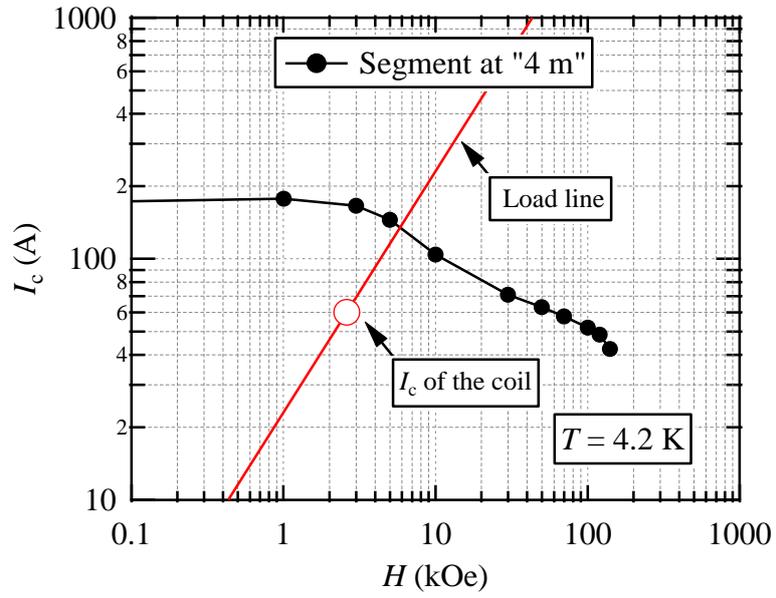

Figure 9. Load line of the $(Ba,Na)Fe_2As_2$ coil and the $I_c$–$H$ curve of a short segment of the wire picked up from the coil at 4.2 K. The $I_c$ of the $(Ba,Na)Fe_2As_2$ coil is indicated by an open circle.